\documentclass[
prd
,byrevtex
,preprint
,showkeys
,showpacs
,amsfonts,amsmath,amssymb
]{revtex4}
\usepackage{bm}
\usepackage[pdftex]{graphicx}
\usepackage{ascmac}
\usepackage{latexsym}
\usepackage{physics}
\begin{document}
\title{Gravity-induced entanglement in optomechanical systems 
}
\author{Akira Matsumura}
\email{matsumura.akira.332@m.kyushu-u.ac.jp}
\author{Kazuhiro Yamamoto}
\email{yamamoto@phys.kyushu-u.ac.jp}
\affiliation{Department of Physics, Kyushu University, Fukuoka, 819-0395, Japan}

\begin{abstract}
We investigate the phenomenon of gravity-induced entanglement in optomechanical systems. 
Assuming photon number conservation and the Newtonian potential expanded up to the quadratic order of the oscillator positions, we exactly solve the dynamics of the optomehcanical systems. Then, we find that the phase difference due to the Newtonian gravity leads to the large entanglement of photons in separated cavities. We clarify the generating mechanism of large gravity-induced entanglements in optomechanical systems in an exact manner. We also determine the characteristic time to generate the maximal entanglement of photons. Finally, by comparing the characteristic time with the decoherence time due to photon leakage, we evaluate the range of the dissipation rate required for testing the gravity-induced entanglement.
\end{abstract}
\keywords{quantum gravity,entanglement, optomechanical system}
\pacs{42.50.-p, 03.65.Ud}
\maketitle

\tableofcontents
\section{Introduction}
One of the most important problems in theoretical physics 
involves establishing
the theory of quantum gravity \cite{Kiefer2006, Woodard2009}. 
The main difficulty is 
that there is no experimental clue regarding quantum gravity and it is even unknown whether the Newtonian gravity is described in the framework of quantum mechanics.
However, the recent advances in quantum sciences might enable experiments to test the
gravitational coupling between small masses in quantum regimes 
(see e.g., Refs.~\cite{Schmole2016, Matsumoto2019,Matsumoto2020}). 
This has motivated theoretical works to explore the quantumness of gravity \cite{Pikovski2012, Albrecht2014, Grossardt2016, Bose2017, Marletto2017, Hall2018, Belenchia2018,CWan,Balushi2018, Carney2019, Carlesso2019, Miao2020, Marshman2020,Krisnanda2020}. 
In these works, tabletop experiments to capture the quantum properties of 
gravity were proposed. 
In particular, Refs. \cite{Bose2017,Marletto2017,Balushi2018} focused on the quantum entanglement induced by gravity.

Quantum entanglement \cite{Nielsen2002, Horodecki2009} is a nonlocal correlation known in quantum mechanics. 
To understand the characteristic feature of quantum entanglement, 
it is important to know the concept of local operations and classical communication (LOCC).
LOCC is the process performed by local observers to exchange only classical information. Mathematically, it is known that quantum entanglement does not increase by LOCC \cite{Nielsen2002, Horodecki2009}. Further, in Ref. \cite{Marletto2017}, 
it was shown that quantum entanglement is not generated by mediating classical systems. The generation of entanglement by gravity implies that the gravitational interactions cannot be described by classical processes. Hence, the detection of gravity-induced entanglement can be a proof of the quantum signature of gravity. 

In the tabletop experiments for testing quantum gravity, the experimental setup with optomechanical systems was proposed \cite{CWan,Balushi2018, Miao2020}. 
An optomechanical system is constructed of an optical cavity and a mechanical mirror ( e.g., \cite{Chen2013, Aspelmeyer2014}). In a sufficiently low temperature, a mirror with approximately $10^{14}$ atoms is described by a quantum harmonic oscillator; it can also exhibit macroscopic superpositions \cite{Marshall2003,Adler2005,Kleckner2008}.
The authors of Refs. \cite{CWan,Balushi2018, Miao2020} investigated the quantum effects of gravity in optomechcanical systems with macroscopic mirrors, 
adopting the Newtonian potential approximated up to the quadratic order of the position
around the equilibrium.

In the present paper, we assume a model similar to that used in
Ref.\cite{Balushi2018} and investigate the generation of photon entanglement by the gravitational interaction between macroscopic mirrors.
Assuming high-finesse mirrors and photon number conservation, we solve the dynamics non-perturbatively for the 
optomechanical couplings in the nonlinear single-photon regime. 
This non-perturbative and nonlinear analysis has not been performed in the previous works \cite{Balushi2018,Miao2020}, which are perturbative or linearized analyses, respectively. 
The author in the Ref.~\cite{CWan} has performed
a non-perturbative analysis, but the quantum 
state has not been given in an exact manner, where
the back action on the mechanical motion due to the gravitational coupling is neglected and the state of photons are assumed to be a pure state periodically. 
We present the exact reduced density matrix for the cavity photons, from which
the full time evolution of entanglement is obtained.
For the dynamics of the model, the oscillator energy depending on the photon states leads to the 
phase difference in the quantum state. We clarify that the phase difference generates the photon entanglement in separated cavities. 
The exact time evolution 
of the photons elucidates the characteristic time of the generation of entanglement.
Furthermore, we discuss the effect of quantum decoherence due to photon leakage.
To maintain quantum entanglement, the decoherence time must be significantly
longer than the time required for generating entanglement.
This condition gives the criterion for the damping 
constant of photons. 

This paper is organized as follows. In Sec.\ref{sec:2}, we introduce the model of cavity optomechanical systems with two mechanical oscillators coupled to each other by gravitational interaction. 
We present the time evolution of the system in an exact manner. 
In Sec.\ref{sec:3}, we examine the behavior of the negativity of photons and the growth time of photon entanglement. We also discuss the phenomenon of photon loss and the criterion required for the dissipation rate to preserve the gravity-induced entanglement. Sec.\ref{sec:4} presents the summary and conclusion. 
In Appendix A, we derive the Hamiltonian of the system. 
In Appendix B, the derivation of Eq.~(\ref{eq:Texp}) is explained in detail. 

\section{Dynamics of optomechanical systems with gravitational interaction}
\label{sec:2}
\begin{figure}[htbp]
  \centering
  \includegraphics[width=0.80\linewidth]{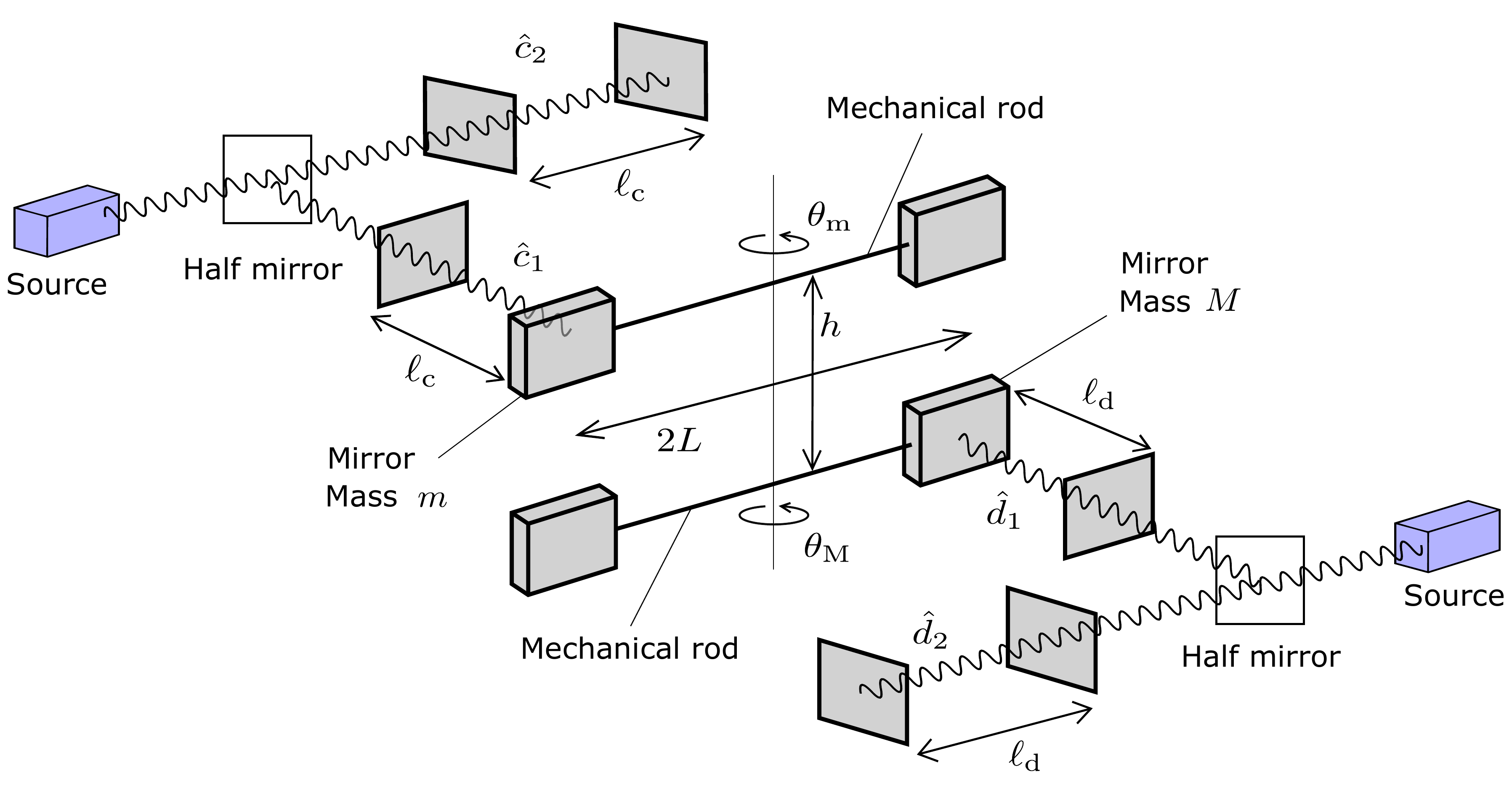}
  \caption{Setup of the cavity optomechanical system, which  
  consists of two mechanical oscillators coupled to the cavity photons \cite{Balushi2018}.}
  \label{fig:setting}
\end{figure}

In this section, we consider the setup of cavity optomechanics to observe the quantum effects of gravitational interactions, which was proposed in Ref. \cite{Balushi2018}. 
Fig.\ref{fig:setting} illustrates the setup consisting of four cavities and two mechanical 
rods suspended from each center. The two rods of length 
$2L$ are separated by a vertical distance 
$h$. Mirrors of mass $m$ and $M$ are fixed at both ends of each rod. 
The cavity photon interacts with the mirror attached to the mechanical rod, and the mirrors at both ends of the rods are coupled with each other by the Newtonian gravity. 
Here, we consider the evolution of the system focusing on the generation of photon entanglement in separated cavities. 
In the above setup, we assume high-reflective (high-finesse) mirrors that preserve the long-term 
evolution of the entanglement. We also assume that the variations in the rods are small and the length of each rod $2L$ is significantly greater than the vertical distance 
$h$, i.e., $\theta_\text{m}, \theta_\text{M}\ll1$ and $2L\gg h$. Then, the total Hamiltonian of the system is defined as follows 
(see Appendix A for the derivation):
\begin{eqnarray}
&&\hat{H}
=\hat{H}_\text{a}+\hat{H}_\text{b}+\hat{H}_\text{a,b} + \hat{H}_\text{c} +\hat{H}_{\text{a},\text{c}_1} +\hat{H}_\text{d}+\hat{H}_{\text{b},\text{d}_1}, 
\label{eq:Htot} 
\\
&&\hat{H}_\text{a}  =  \frac{\hbar \omega_\text{a}}{2} (\hat{p}^2_\text{a} + \hat{q}^2_\text{a}), 
\quad 
\hat{H}_\text{b} =  \frac{\hbar \omega_\text{b}}{2} (\hat{p}^2_\text{b} + \hat{q}^2_\text{b}), 
\label{eq:Hfreeab}
\\
&&\hat{H}_\text{c} =\hbar \omega_\text{c} (\hat{c}^\dagger_1 \hat{c}_1+\hat{c}^\dagger_2 \hat{c}_2), 
\quad
\hat{H}_\text{d} =\hbar \omega_\text{d} (\hat{d}^\dagger_1 \hat{d}_1+\hat{d}^\dagger_2 \hat{d}_2), 
\label{eq:Hfreecd} 
\\
&&\hat{H}_{\text{a},\text{c}_1} 
=-\lambda_\text{m} \hbar \omega_\text{a} \hat{c}^\dagger_1 \hat{c}_1 \hat{q}_\text{a}, 
\quad 
\hat{H}_{\text{b},\text{d}_1}
=-\lambda_M \hbar \omega_\text{b} \hat{d}^\dagger_1 \hat{d}_1 \hat{q}_\text{b}, 
\quad
\hat{H}_\text{a,b} = - \hbar \gamma \hat{q}_\text{a} \hat{q}_\text{b}, 
\label{eq:Hint}
\end{eqnarray}
where 
$\hat{q}_\text{a}$,
$\hat{p}_\text{a}$,
$\hat{q}_\text{b}$, and 
$\hat{p}_\text{b}$ are the dimensionless canonical variables of the mechanical oscillators (the oscillating rods). These variables satisfy the following canonical commutation relations.
\begin{equation}
[\hat{q}_\text{a}, \hat{q}_\text{a}] = [\hat{p}_\text{a}, \hat{p}_\text{a}]=0, 
\quad
[\hat{q}_\text{a}, \hat{p}_\text{a}]=i,
\label{eq:qapa}
\end{equation}
\begin{equation}
[\hat{q}_\text{b}, \hat{q}_\text{b}] = [\hat{p}_\text{a}, \hat{p}_\text{a}]=0, 
\quad
[\hat{q}_\text{b}, \hat{p}_\text{b}]=i. 
\label{eq:qbpb}
\end{equation}
Here, the frequencies 
$\omega_\text{a}$ and 
$\omega_\text{b}$ are defined as
\begin{equation}
\omega^2_\text{a}=\Omega^2_\text{a}+\frac{GM}{h^3} , 
\quad 
\omega^2_\text{b}=\Omega^2_\text{b}+\frac{Gm}{h^3}, 
\label{eq:omegaab_appendix}
\end{equation}
where 
$\Omega_\text{a}$ and 
$\Omega_\text{b}$ are the natural frequencies of each oscillator in the horizontal plane, and
$G$ is the Newtonian constant of gravitation. 
The annihilation (creation) operators of 
the photons with frequencies  
$\omega_\text{c}$ and 
$\omega_\text{d}$ in the cavities are denoted by 
$\hat{c}_1$($\hat{c}_1^\dagger$), 
$\hat{c}_2$($\hat{c}_2^\dagger$), 
$\hat{d}_1$($\hat{d}_1^\dagger$), and 
$\hat{d}_2$($\hat{d}_2^\dagger$), where
$\hat{c}_1$($\hat{c}_1^\dagger$) and $\hat{d}_1$($\hat{d}_1^\dagger$) 
are the photon operators that interact with 
each mirror attached to the rods, while 
$\hat{c}_2$($\hat{c}_2^\dagger$) and $\hat{d}_2$($\hat{d}_2^\dagger)$
are those trapped in other cavities without any coupling to the oscillators. 
The coupling constants $\lambda_\text{m}$, 
$\lambda_\text{M}$, and $\gamma$ of Eq. \eqref{eq:Hint} are   
\begin{equation}
\lambda_\text{m} 
=\frac{\omega_\text{c}}{\omega_\text{a} \ell_\text{c}} \sqrt{\frac{\hbar}{2m \omega_\text{a}}},
\quad 
\lambda_\text{M}
= \frac{\omega_\text{d}}{\omega_\text{b} \ell_\text{d}} \sqrt{\frac{\hbar}{2M \omega_\text{b}}}, 
\quad
\gamma
= \frac{G}{h^3} \sqrt{\frac{Mm}{\omega_\text{a} \omega_\text{b}}}, 
\label{eq:lambdagamma}
\end{equation}
where
$\ell_\text{c}$ and 
$\ell_\text{d}$ are the lengths of the cavities with the photons described by
$\hat{c}_1$($\hat{c}_1^\dagger$), $\hat{c}_2$($\hat{c}_2^\dagger)$ and 
$\hat{d}_1$($\hat{d}_1^\dagger$), $\hat{d}_2$($\hat{d}_2^\dagger)$, 
respectively. Introducing the correction parameters associated with the gravity, 
\begin{equation}
g_\text{a}= \frac{GM}{\Omega^2_\text{a} h^3}, 
\quad
g_\text{b}= \frac{Gm}{\Omega^2_\text{b} h^3},
\label{eq:g}
\end{equation}
we can rewrite 
$\omega_\text{a}, \, \omega_\text{b}$,
$\lambda_\text{m}, \, \lambda_\text{M},$ and $ \gamma$ as 
\begin{equation}
\omega^2_\text{a}=(1+g_\text{a}) \Omega^2_\text{a} , 
\quad 
\omega^2_\text{b}=(1+g_\text{b}) \Omega^2_\text{b}, 
\label{eq:omegaab2}
\end{equation}
and 
\begin{equation}
\lambda_\text{m} 
=\frac{\Lambda_\text{m}}{(1+g_\text{a})^{3/4}},
\quad 
\lambda_\text{M}
= \frac{\Lambda_\text{M}}{(1+g_\text{b})^{3/4}}, 
\quad
\gamma
=\frac{\sqrt{g_\text{a}  g_\text{b} \Omega_\text{a} \Omega_\text{b}}}{(1+g_\text{a})^{1/4} (1+g_\text{b})^{1/4}},
\label{eq:lambdagamma2}
\end{equation}
where 
$\Lambda_\text{m}$ and 
$\Lambda_\text{M}$ are the standard optomechcanical couplings;
\begin{equation}
\Lambda_\text{m} 
=\frac{\omega_\text{c}}{\Omega_\text{a} \ell_\text{c}} \sqrt{\frac{\hbar}{2m \Omega_\text{a}}},
\quad 
\Lambda_\text{M} 
=\frac{\omega_\text{d}}{\Omega_\text{b} \ell_\text{d}} \sqrt{\frac{\hbar}{2M \Omega_\text{b}}}.
\label{eq:Lambda}
\end{equation}
In Ref.~\cite{Balushi2018}, the time evolution generated by the total Hamiltonian \eqref{eq:Htot} was given by the perturbative expansion of the optomechanical couplings 
$\lambda_\text{m}$ and $\lambda_\text{M}$. 
In Ref.~\cite{Miao2020}, the authors adopted the linearized theory 
for the canonical variables of the mechanical oscillators and cavity fields. 

However, in the present paper, we solve the nonlinear dynamics with a
non-perturbative method using photon number conservation. 
Accordingly, we consider the interaction picture defined by 
$|\Psi_\text{I} (t) \rangle = e^{i \hat{H}_0 t/\hbar} |\Psi(t) \rangle$
with $\hat{H}_0 = \hat{H}_\text{a}+\hat{H}_\text{b}+\hat{H}_\text{a,b} + \hat{H}_\text{c} +\hat{H}_\text{d}$, 
where it should be noted that the total Hamiltonian is written as
$\hat H=\hat H_0+\hat V$ with
$\hat{V}=\hat{H}_{\text{a},\text{c}_1}+\hat{H}_{\text{b},\text{d}_1}$. 
The evolved state in the interaction picture is given by
\begin{equation}
|\Psi_\text{I} (t) \rangle 
= \text{T} \exp \Bigl[ -\frac{i}{\hbar} \int ^t_0 dt' \hat{V}_\text{I} (t') \Bigr] 
|\Psi_\text{I}(0) \rangle, 
\label{eq:sol}
\end{equation}
where 
$\hat{V}_\text{I} (t)=e^{i\hat{H}_0 t/ \hbar} \hat{V} e^{-i\hat{H}_0 t/ \hbar}$ and T is the time order product. 
For convenience, introducing the operators in the vector notation,
\begin{equation}
\hat{\bm{\eta}} 
= 
[ \hat{q}_\text{a}, \hat{q}_\text{b}, \hat{p}_\text{a}, \hat{p}_\text{b} ]^{\text{T}}, 
\quad
\hat{\bm{j}}
= 
[\lambda_\text{m} \omega_\text{a} \hat{c}^\dagger_\text{1}\hat{c}_\text{1},  \lambda_\text{M} \omega_\text{b} \hat{d}^\dagger_\text{1}\hat{d}_\text{1}, 0, 0]^\text{T}, 
\label{eq:etah}
\end{equation}
we can write
$\hat{V}= -\hbar \, \hat{\bm{j}}^\text{T} \hat{\bm{\eta}}$; then,  $\hat{V}_\text{I}(t)/\hbar$ is given by 
\begin{equation}
\hat{V}_\text{I}(t)/\hbar
= -\hat{\bm{j}}^\text{T}_\text{I} (t) \hat{\bm{\eta}}_\text{I} (t)
= -\hat{\bm{j}}^\text{T} \hat{\bm{\eta}}_\text{I} (t),
\label{eq:VI}
\end{equation}
where
$ \hat{\bm{j}}_\text{I} (t)= \hat{\bm{j}}$. 
In the interaction picture, the canonical operator 
$\hat{\bm{\eta}}_\text{I} (t) $ of the oscillators is
\begin{equation}
\hat{\bm{\eta}}_\text{I}(t)
=
e^{\frac{it}{\hbar} (\hat{H}_\text{a} + \hat{H}_\text{b} +\hat{H}_\text{a,b})}\,
\hat{\bm{\eta}} \,
e^{-\frac{it}{\hbar} (\hat{H}_\text{a} + \hat{H}_\text{b} +\hat{H}_\text{a,b})}
=
e^{\frac{it}{2}\hat{\bm{\eta}}^\text{T} \mathcal{H} \hat{\bm{\eta}}} \,
\hat{\bm{\eta}} \, 
e^{-\frac{it}{2}\hat{\bm{\eta}}^\text{T} \mathcal{H} \hat{\bm{\eta}}}
= e^{t\Omega \mathcal{H}} \hat{\bm{\eta}}, 
\label{eq:etaeq}
\end{equation}
where the Hamiltonian 
$\hat{H}_\text{a}+\hat{H}_\text{b}+\hat{H}_\text{a,b}$ for the vector notation was rewritten as 
\begin{equation}
\hat{H}_\text{a}+\hat{H}_\text{b}+\hat{H}_\text{a,b}=\frac{\hbar}{2} \hat{\bm{\eta}}^\text{T} \mathcal{H} \hat{\bm{\eta}},
\quad
\mathcal{H} =
\begin{bmatrix}
\omega_\text{a} && -\gamma  && 0 && 0 \\
-\gamma && \omega_\text{b} && 0 && 0 \\
0 && 0 && \omega_\text{a} && 0 \\
0 && 0 && 0 && \omega_\text{b}
\end{bmatrix},
\label{eq:mathH}
\end{equation}
additionally, we used the Backer-Campbell-Hausdorff formula 
$e^{\hat{A}}\hat{B} e^{-\hat{A}}
=\sum_{n} [\hat{A},[\hat{A},[\cdots,[\hat{A},\hat{B}]]\cdots]/n!$ 
and the canonical commutation relations 
$[\hat{\eta}^j , \hat{\eta}^k ] =i \Omega^{jk}$ with the symplectic 
matrix 
$\Omega$. Using the formula of $\hat{\eta}_\text{I} (t)$ in Eq.~\eqref{eq:etaeq} and considering that the interaction 
$\hat{V}_\text{I} (t)$ in Eq.~\eqref{eq:VI} does not change the photon number, we obtain the equation
$[\hat{V}_\text{I}(t_1),[\hat{V}_\text{I}(t_2), \hat{V}_\text{I}(t_3)]]=0$.  
From this equation, we can compute the time order product in Eq.~\eqref{eq:sol} as
\begin{equation}
\text{T} \exp \Bigl[ -\frac{i}{\hbar} \int ^t_0 dt' \hat{V}_\text{I} (t') \Bigr]
= 
\exp 
\Bigl[
i \int^t_0 dt' \, \hat{\bm{j}}^\text{T} \hat{\bm{\eta}}_\text{I} (t') 
-\frac{i}{2} \int^t_0 dt' \int^{t}_0 dt'' \, \hat{\bm{j}}^\text{T} {\cal G}(t', t'')\hat{\bm{j}}  
\Bigr],
\label{eq:Texp}
\end{equation}
where matrix 
${\cal G}(t', t'')$ is the retarded Green's function of the linear differential equation 
$d \hat{\bm{\eta}}_\text{I} /dt =\Omega \mathcal{H} \hat{\bm{\eta}}_\text{I}$ with the components 
${\cal G}^{jk} (t', t'')
= 
-i \, [ \hat{\eta}^j_\text{I} (t'), \hat{\eta}^k_\text{I} (t'') ] \theta(t'-t'')$. In Appendix~\ref{sec:B}, Eq. \eqref{eq:Texp} is derived. The first term of the exponent in Eq.~\eqref{eq:Texp} represents the interaction between the photons and oscillators. The second term corresponds to the photon-photon interaction mediated by the oscillators. We expect that the quantum entanglement of photons is generated by the second term. Substituting the formula of 
$\hat{\bm{\eta}}_\text{I}$ (Eq.(\ref{eq:etaeq})) into Eq.~\eqref{eq:Texp} and performing time integration, we find the evolved state of the total system (photons and oscillators) in the interaction picture as follows: 
\begin{equation}
| \Psi_\text{I} (t) \rangle
= 
e^{i \, \hat{\bm{j}}^\text{T} 
\mathcal{H}^{-1} \Omega [ \mathbb{I}-e^{t\Omega \mathcal{H}}] \hat{\bm{\eta}} 
+\frac{i}{2} \hat{\bm{j}}^\text{T} 
\bigl(
t\mathcal{H}^{-1} -[ \mathbb{I}-e^{t\Omega \mathcal{H}}] \mathcal{H}^{-1} \Omega \mathcal{H}^{-1}
\bigr) \hat{\bm{j}}} \, 
|\Psi_\text{I}(0) \rangle.
\label{eq:PsiIt}
\end{equation}
Here, the term  $\hat{\bm{j}}^\text{T} 
\mathcal{H}^{-1} \hat{\bm{j}} /2$
in the exponent of Eq.~\eqref{eq:PsiIt} is proportional to the minimum energy of the oscillators coupled with the photons (up to the zero-point energy), which depends on the photon state. 
Namely, the Hamiltonian of the oscillators is written as 
\begin{eqnarray}
\hat{H}_\text{osc}/\hbar ={1\over 2}\hat{\bm {\eta}}^\text{T} {\cal H}\hat{\bm{\eta}}-\hat{\bm{j}}^\text{T} \hat{{\bm \eta}},
\end{eqnarray}
where 
$\hat{\bm{\eta}}$ and 
$\hat{\bm{j}}$ were introduced in Eq.\eqref{eq:etah}, and 
$\cal H$ was defined in Eq.\eqref{eq:mathH}. We note that $\hat{H}_\text{osc}/\hbar$ is rewritten as 
\begin{eqnarray}
\hat{H}_\text{osc}/\hbar ={1\over 2}(\hat{\bm {\eta}}-\mathcal{H}^{-1}\hat{\bm j})^T 
{\cal H}(\hat{{\bm \eta}}-\mathcal{H}^{-1}\hat{\bm j}) -{1\over 2} \hat{\bm {j} }^\text{T} \mathcal{H}^{-1}\hat{\bm {j}}  
\label{twotwo}
\end{eqnarray}
Therefore, the minimum energy of $\hat{H}_\text{osc}$ is given by the zero point energy and 
$- \hbar \hat{\bm{ j}}^\text{T} \mathcal{H}^{-1}\hat{\bm {j}}/2$. 
The latter contribution 
changes the phase of photon states depending on the number of photons. 
In the next section, we will show that the phase difference induced by the minimum energy 
leads to photon entanglement in separated cavities.

\section{Gravity-induced entanglement between photons}
\label{sec:3}

In this section, we focus on the gravity-induced entanglement of photons
and derive the typical time scale for its generation. 
The modes $c_2$ and $d_2$, which looks decoupled from the system apparently,  play an important role to generate quantum superposition in photons as well as quantum entanglement. 
For simplicity, the photons in the cavities are initially assumed to be in the 
single-photon superposed state 
\begin{equation}
|\phi \rangle_\text{c,d} 
= 
\frac{1}{\sqrt{2}} \Bigl( | 0,1 \rangle_\text{c} + | 1,0 \rangle_\text{c} \Bigr)  
\otimes 
\frac{1}{\sqrt{2}} \Bigl( | 0,1 \rangle_\text{d} + | 1,0 \rangle_\text{d} \Bigr),
\label{eq:inphoton}
\end{equation}
where
$| 0,1 \rangle_\text{c}=\hat{c}^\dagger_{2} |0,0 \rangle_\text{c}$, 
$| 1,0 \rangle_\text{c}=\hat{c}^\dagger_{1} |0,0 \rangle_\text{c}$, 
$| 0,1 \rangle_\text{d}=\hat{d}^\dagger_{2} |0,0 \rangle_\text{d}$, and
$| 1,0 \rangle_\text{d}=\hat{d}^\dagger_{1} |0,0 \rangle_\text{d}$.
Quantum state of the oscillators evolves under the
influence of the gravity, which depends on the state
of photons in the superposition state. 
We focus on the quantum entanglement of photons
of the states spanned by $\{
| 0,1 \rangle_\text{c}\otimes| 0,1 \rangle_\text{d},~
| 0,1 \rangle_\text{c}\otimes| 1,0 \rangle_\text{d},~
| 1,0 \rangle_\text{c}\otimes| 0,1 \rangle_\text{d},~
| 1,0 \rangle_\text{c}\otimes| 1,0 \rangle_\text{d}\}$.

Furthermore, we assume that the initial state of the oscillators is the coherent state
\begin{equation}
| \bm{\xi} \rangle_\text{a,b} = \hat{W}(\bm{\xi}) |0 \rangle_\text{a,b}, \label{eq:xi}
\end{equation}
where 
$\hat{W} (\bm{\xi})= \exp [ i \bm{\xi}^\text{T} \Omega \hat{\bm{\eta}} ]  \, ( \bm{\xi} \in \mathbb{R}^4) $ is the Weyl operator. Here we have assumed the initial coherent state, however, we find that the entanglement of photons does not depend on the parameter $\bm{\xi}$, which will be explained below Eq.\eqref{eq:explicitN}.
In a realistic situation, the oscillators may be in thermal states, which leads to thermal noises and the degradation of entanglement. In the present paper, we do not discuss the noise effect and only focus on the phenomenon of 
entanglement generation.

From the solution of Eq.\eqref{eq:PsiIt} in the interaction picture, the evolved state 
$|\Psi(t) \rangle=e^{-i\hat{H}_0 t/\hbar} |\Psi_\text{I} (t) \rangle$ with 
$\hat{H}_0 = \hat{H}_\text{c}+\hat{H}_\text{d}+\hat{H}_\text{a}+\hat{H}_\text{b}+\hat{H}_\text{a,b}$ in the Schr\"odinger picture is given as
\begin{align}
| \Psi(t) \rangle 
&
=
e^{-i\hat{H}_0 t/ \hbar} \,
e^{i \, \hat{\bm{j}}^\text{T} 
\mathcal{H}^{-1} \Omega [ \mathbb{I}-e^{t\Omega \mathcal{H}}] \hat{\bm{\eta}} 
+\frac{i}{2} \hat{\bm{j}}^\text{T} 
\bigl(
t\mathcal{H}^{-1} -[ \mathbb{I}-e^{t\Omega \mathcal{H}}] \mathcal{H}^{-1} \Omega \mathcal{H}^{-1}
\bigr) \hat{\bm{j}}} \, 
|\phi \rangle_\text{c,d}  \otimes | \bm{\xi} \rangle_\text{a,b}
\nonumber 
\\
&
=
\frac{1}{2}e^{-i(\omega_\text{c}+\omega_\text{d})t} 
\sum^1_{n,m=0}
|n,1-n \rangle_\text{c} |m,1-m \rangle_\text{d}
 \otimes e^{-i (\hat{H}_\text{a}+\hat{H}_\text{b}+\hat{H}_\text{a,b}) t/\hbar} | \psi_{nm} \rangle,
\label{eq:Psit2}
\end{align}
where the initial state of the total system is 
$|\Psi_\text{I} (0) \rangle 
=|\phi \rangle_\text{c,d}  \otimes | \bm{\xi} \rangle_\text{a,b}$ and the state of the mechanical oscillators 
$|\psi_{nm}\rangle$ is
\begin{equation}
| \psi_{nm} \rangle 
=
e^{\frac{i}{2} \bm{j}^\text{T}_{nm} 
\bigl(
t\mathcal{H}^{-1} -[ \mathbb{I}-e^{t\Omega \mathcal{H}}] \mathcal{H}^{-1} \Omega \mathcal{H}^{-1}
\bigr) \bm{j}_{nm} 
 + i \, \bm{j}^\text{T}_{nm} 
\mathcal{H}^{-1} \Omega [ \mathbb{I}-e^{t\Omega \mathcal{H}}] \hat{\bm{\eta}}} 
| \xi \rangle_\text{a,b}.
\label{eq:psinm}
\end{equation}
The vector 
$\bm{j}_{nm}=[ \lambda_\text{m} \omega_\text{a} n, \lambda_\text{M} \omega_\text{b} m, 0,0]^\text{T}$ is the eigenvalue of $\hat{\bm{j}}$, 
where $n$ and $m$ are the eigenvalues of the number operators 
$\hat{c}^\dagger_1 \hat{c}_1$ and $\hat{d}^\dagger_1\hat{d}_1$, respectively.

Tracing over the degree of freedom of the mechanical oscillators, we obtain the reduced density operator,
\begin{align}
\rho_\text{c,d} (t) 
&
=
\text{Tr}_\text{a,b} \Bigl[|\Psi(t) \rangle \langle \Psi(t)| \Bigr] 
\nonumber 
\\
&
=\frac{1}{4} 
\begin{bmatrix}
1 
&& \langle \psi_{01} | \psi_{00}\rangle 
&& \langle \psi_{10} | \psi_{00} \rangle 
&& \langle \psi_{11} | \psi_{00} \rangle \\
*
&& 1
&& \langle \psi_{10} | \psi_{01} \rangle 
&& \langle \psi_{11} | \psi_{01} \rangle \\
*
&& *
&& 1
&& \langle \psi_{11} | \psi_{10}\rangle \\
*
&& *
&& *
&& 1
\end{bmatrix}
\label{eq:rhocd},
\end{align}
where the order of the basis is 
$\{ |0,1 \rangle_\text{c} |0,1 \rangle_\text{d}, \,
|0,1 \rangle_\text{c} |1,0 \rangle_\text{d}, \,
|1,0 \rangle_\text{c} |0,1 \rangle_\text{d}, \,
|1,0 \rangle_\text{c} |1,0 \rangle_\text{d}   \} $, and 
$*$ is determined by considering that  
$\rho_\text{c,d} (t)$ is the self-adjoint operator.
The off-diagonal components, which are given by the inner product 
$\langle \psi_{nm}| \psi_{n'm'} \rangle$,
characterize the entanglement between the degree of freedoms of photons labeled by c and d. Explicitly, the inner product is
\begin{equation}
\langle \psi_{nm} | \psi_{n'm'}\rangle 
= 
e^{ i \, \bm{\xi}^\text{T} \Omega  (\mathbb{I}-e^{-t\Omega \mathcal{H}}) \mathcal{H}^{-1} \bm{j}_{-}+ \frac{i}{2} \bm{j}^\text{T}_{+} \bigl[ t \mathcal{H}^{-1} + (\mathbb{I}-e^{-t\Omega\mathcal{H}} ) \mathcal{H}^{-1} \Omega \mathcal{H}^{-1} \bigr] 
\bm{j}_{-} -\frac{1}{4} \left| [\mathbb{I}-e^{-t\Omega \mathcal{H} }]  \mathcal{H}^{-1}\bm{j}_{-}\right|^2},
\label{eq:inn2}
\end{equation} 
where we defined the vector
$\bm{j}_{\pm}= \bm{j}_{nm} \pm  \bm{j}_{n'm'}$ and 
$|\bm{v}|^2=\bm{v}^\text{T} \bm{v}$ for a vector 
$\bm{v}$. Eq. \eqref{eq:inn2} can be obtained from the inner product of two coherent states.

To evaluate the quantum entanglement, we introduce the negativity \cite{Vidal2002}, which is a useful entanglement measure based on the positive partial transpose criterion \cite{Peres1996, Horodecki1996} . The negativity is given by the sum of the negative eigenvalues 
($\lambda_i <0$) of a partial transposed density matrix $\rho_\text{AB}^{\text{T}_\text{A}}$,   
\begin{equation}
\mathcal{N}=\sum_{\lambda_{i}<0} |\lambda_{i}|.
\end{equation}
If the negativity does not
vanish (there are some negative eigenvalues of
$\rho^{\text{T}_\text{A}}_\text{AB}$), then the state is entangled \cite{Peres1996, Horodecki1996, Vidal2002}. The converse is specifically true when the Hilbert space
$\mathcal{H}_\text{A} \otimes \mathcal{H}_\text{B}$ is
$\mathbb{C}^{2} \otimes \mathbb{C}^{2}$ or
$\mathbb{C}^{2} \otimes \mathbb{C}^{3}$ \cite{Horodecki1996}. Since the system of the photons of interest effectively corresponds to a two-qubit system ($\mathbb{C}^{2} \otimes \mathbb{C}^{2}$), the negativity of the photons has a nonzero value if and only if the photon state is entangled. 

Using the formula of the inner product (Eq.\eqref{eq:inn2}), we find the four eigenvalues of 
the transposed matrix $\rho^{\text{T}_\text{c}}_\text{c,d}$: 
\begin{align}
&
\frac{1}{4} 
\Bigl(
1-e^{-A-B} \cosh[C]
\pm \sqrt{
(e^{-A}-e^{-B} )^2 
+4e^{-A-B} \sin^2 [\Theta/2] 
+e^{-2A-2B} \sinh^2[C]}
\Bigr), 
\nonumber
\\
&
\frac{1}{4} 
\Bigl(
1+e^{-A-B} \cosh[C]
\pm \sqrt{
(e^{-A}-e^{-B} )^2 
+4e^{-A-B} \sin^2 [\Theta/2] 
+e^{-2A-2B} \sinh^2[C]}
\Bigr),
\label{eq:egnvalue}
\end{align}
where
\begin{eqnarray}
&&\Theta
=
\bm{j}^\text{T}_{10} 
\Bigl[
t\mathcal{H}^{-1}
+\frac{1}{2} (e^{t\Omega \mathcal{H}}-e^{-t\Omega \mathcal{H}})\mathcal{H}^{-1} \Omega \mathcal{H}^{-1}
\Bigr]
\bm{j}_{01},
\label{eq:Theta}
\\
&&A
=\frac{1}{4} |(\mathbb{I}-e^{-t\Omega \mathcal{H}})\mathcal{H}^{-1} \bm{j}_{10}|^2, 
\label{eq:AB}
\\
&&B
=\frac{1}{4} |(\mathbb{I}-e^{-t\Omega \mathcal{H}})\mathcal{H}^{-1} \bm{j}_{01}|^2,
\label{eq:ABQ}
\\
&&C
=\frac{1}{2}  \bm{j}_{10}^\text{T} [(\mathbb{I}-e^{-t\Omega \mathcal{H}})\mathcal{H}^{-1}]^\text{T}(\mathbb{I}-e^{-t\Omega \mathcal{H}})\mathcal{H}^{-1} \bm{j}_{01}.
\label{eq:C}
\end{eqnarray}
Since the absolute value of the inner product 
$\langle \psi_{nm}| \psi_{n'm'} \rangle$ for 
$n,m,n',m'=0,1$ is smaller than or equal to unity, the inequality 
$A+B \pm C \geq 0$ is satisfied. From this inequality, we find that one of the four eigenvalues can be negative, and obtain the negativity as follows :  
\begin{equation}
\mathcal{N}
=
\max \Bigl[
-\frac{1}{4}  
\Bigl(
1-e^{-A-B} \cosh[C]
-\sqrt{
(e^{-A}-e^{-B} )^2 
+4e^{-A-B} \sin^2 [\Theta/2] 
+e^{-2A-2B} \sinh^2[C]}
\Bigr), 0
\Bigr].
\label{eq:explicitN}
\end{equation} 
The negativity does not depend on the initial displacement 
$\bm{\xi}$. This is because the Newtonian potential generates the entanglement for the spatial superposed state of the oscillators, and such a superposition dose not depend on 
$\bm{\xi}$. This also corresponds to  the fact that the 
$\bm{\xi}$ dependence can be eliminated by local unitary transformations, which does not change the quantum entanglement. 

Next, we examine the time dependence appearing in the formula of negativity. In the expressions of 
$A, B$, and 
$C$ (Eqs.~\eqref{eq:AB}, \eqref{eq:ABQ} and \eqref{eq:C}, respectively), the time dependence is determined
through the matrix $e^{-t\Omega \mathcal{H}}$. The eigenvalues of $\Omega \mathcal{H}$ are 
$\pm i \omega_{+}$ and 
$\pm i \omega_{-}$, where
\begin{equation}
\omega_{\pm} 
=
\sqrt{ \frac{\omega^2_\text{a}+\omega^2_\text{b}
\pm 
\sqrt{(\omega^2_\text{a}-\omega^2_\text{b})^2
+4\gamma^2 \omega_\text{a} \omega_\text{b}}}{2}}.
\label{eq:fast}
\end{equation}
For a large $t$, $\Theta$ is approximated as
$\Theta\sim \omega_\text{s}t$, where $\omega_\text{s} $ is given by
\begin{equation}
\omega_\text{s} 
= \bm{j}^\text{T}_{10} \mathcal{H}^{-1} \bm{j}_{01} 
= \gamma \lambda_\text{m} \lambda_\text{M} 
\Bigl[1- \frac{\gamma^2}{\omega_\text{a} \omega_\text{b}} \Bigr]^{-1}.
\label{eq:slow}
\end{equation}
Hence, the negativity typically varies with the three time scales, i.e.,  $1/\omega_{+}$, 
$1/\omega_{-}$, and $1/\omega_\text{s}$. 
We note that $\omega_s \geq 0$ because the couplings 
$\lambda_\text{m}$,
$\lambda_\text{M}$, and 
$\gamma$, defined in Eq.~\eqref{eq:lambdagamma},
are positive and the inequality $\omega_\text{a}\omega_\text{b}\geq \gamma^2 $ is satisfied from Eqs.~\eqref{eq:omegaab2} and \eqref{eq:lambdagamma2}.

\begin{figure}[t]
   \centering
   \includegraphics[width=0.495\linewidth]{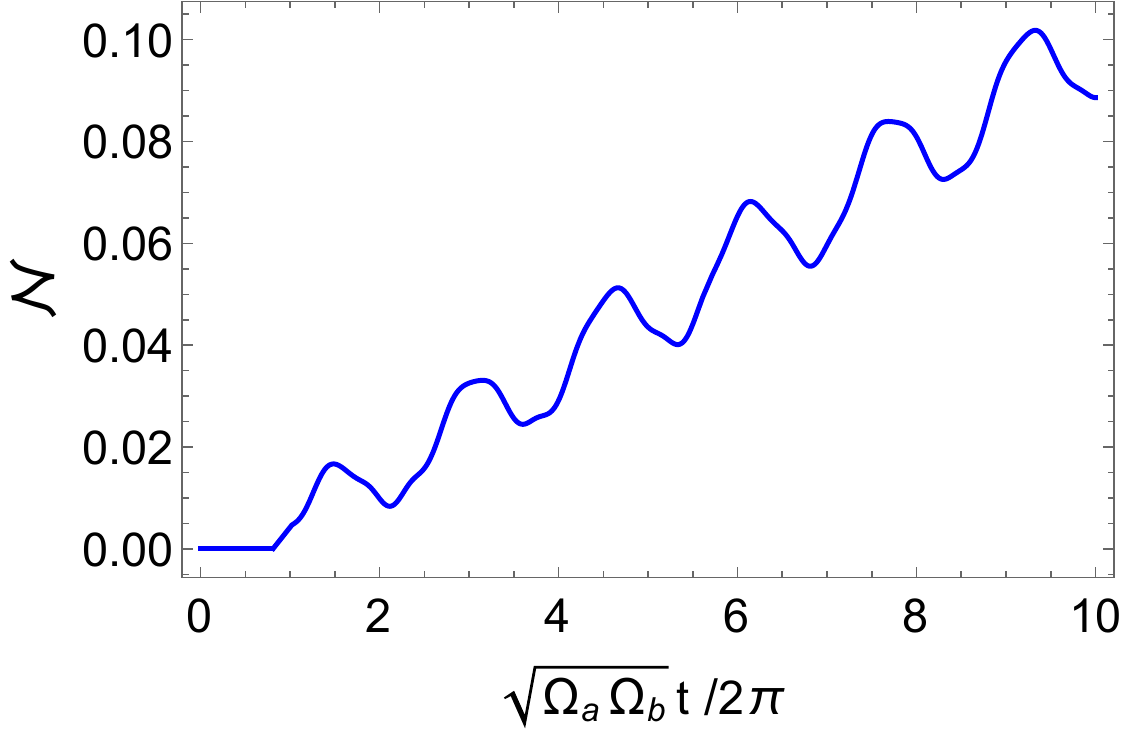}
   \includegraphics[width=0.495\linewidth]{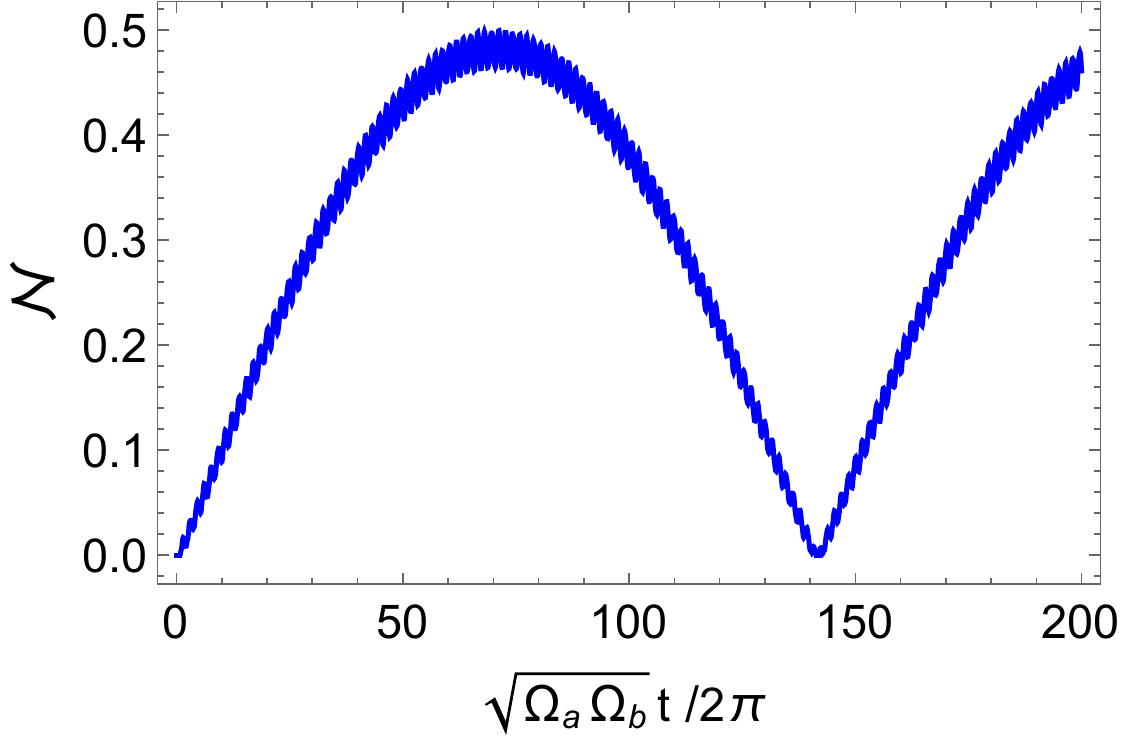}
   \caption{Left panel: behavior of the negativity for $0 \leq \sqrt{\Omega_\text{a} \Omega_\text{b}} t/2\pi \leq 10$. Right panel: behavior of the negativity for $0 \leq \sqrt{\Omega_\text{a} \Omega_\text{b}}  t/2\pi \leq 200$. For each plot, the parameters are fixed as $\Omega_\text{a}/ \Omega_\text{b}=3.0,\, \Lambda_\text{m}=0.1, \, \Lambda_\text{M}=0.3,\, g_\text{a}=0.5$, and $g_\text{b}=0.4$. }
    \label{fig:ng}
\end{figure}

We demonstrate the behavior of negativity $\mathcal{N}$ in time. The left panel of Fig. \ref{fig:ng} 
shows the time evolution of  
$\mathcal{N}$ for 
$0 \leq \sqrt{\Omega_\text{a} \Omega_\text{b}} t/2\pi \leq 10$, while the right panel 
shows that for
$0 \leq \sqrt{\Omega_\text{a} \Omega_\text{b}} t/2\pi \leq 200$. 
Here, the optomechanical couplings without the Newtonian gravity are fixed as 
$\Lambda_\text{m}=0.1$ and 
$\Lambda_\text{M}=0.3$. The ratio of the natural frequencies is 
$\Omega_\text{a}/\Omega_\text{b}=3.0$, and the parameters related to the gravity are $g_\text{a}=0.5$ and $g_\text{b}=0.4$. 
In the early stages of time evolution (left panel of Fig. \ref{fig:ng} ), 
$\mathcal{N}$ grows in proportion to time $t$. This feature was first observed 
in Ref. \cite{Balushi2018}. 
However, this growth stops and $\mathcal{N}$ reaches a maximum value. For the parameters chosen in Fig. \ref{fig:ng}, 
the typical timescales of $\mathcal{N}$ are 
$\sqrt{\Omega_\text{a} \Omega_\text{b}}/2\pi \omega_{\pm} \sim O(1)$ and
$\sqrt{\Omega_\text{a} \Omega_\text{b}}/2\pi \omega_\text{s} \sim O(10)$. 

Let us consider the long-term behavior of $\mathcal{N}$. 
We focus on the minimum energy of the oscillators coupled to the photons,  which is determined by $-\hbar \hat{\bm{j}}^\text{T} \mathcal{H}^{-1} \hat{\bm{j}}/2$ and was explicitly shown in Eq. \eqref{twotwo}. 
Using the definitions 
$\hat{\bm{j}}$ and $\mathcal{H}$ defined in Eqs.\eqref{eq:etah} and \eqref{eq:mathH}, respectively, we find that 
 $-\hbar \hat{\bm{j}}^\text{T} \mathcal{H}^{-1} \hat{\bm{j}}/2$ includes the term 
$\hbar \omega_\text{s} \, \hat{c}^\dagger_1 \hat{c}_\text{1}  
\hat{d}^\dagger_1 \hat{d}_1$, where $\omega_\text{s}$ is the frequency introduced in Eq.\eqref{eq:slow}. This term gives the phase difference and generates the entanglement for photon states as follows:
\begin{align}
&
e^{i \omega_\text{s} t \, 
\hat{c}^\dagger_1 \hat{c}_\text{1} 
\hat{d}^\dagger_1 \hat{d}_1} \frac{1}{\sqrt{2}} \Bigl( | 0,1 \rangle_\text{c} + | 1,0 \rangle_\text{c} \Bigr)  \otimes 
\frac{1}{\sqrt{2}} \Bigl( | 0,1 \rangle_\text{d} + | 1,0 \rangle_\text{d} \Bigr)
\nonumber 
\\
&
\qquad
= \frac{1}{2} \bigl( 
| 0,1 \rangle_\text{c}
+
| 1,0 \rangle_\text{c} 
\bigr) 
\otimes 
| 0,1 \rangle_\text{d} 
+ 
\frac{1}{2}
\bigl( 
| 0,1 \rangle_\text{c}
+
e^{i\omega_\text{s} t} 
| 1,0 \rangle_\text{c}
\bigr)
\otimes 
| 1,0 \rangle_\text{d}.
\label{eq:entgene}
\end{align}
This state is entangled excepting at the moments
$\omega_\text{s}t=2n\pi \,(n \in \mathbb{N})$, while the maximal entangled state appears at $\omega_\text{s}t=(2n-1)\pi$. This explains the long-term behaviour in the right panel of Fig. \ref{fig:ng}. We note that 
the fast oscillations in the left panel of Fig. \ref{fig:ng} is explained by the short time scale described by
$1/\omega_\pm$.
The time scale $1/\omega_\text{s}$ is important for experimental tests to generate a large entanglement by the gravity. 

We should compare $ 1/\omega_\text{s}$ with the decoherence time for  the optomechanical 
cavities to discuss the generation of large entanglements.   
Quantum decoherence occurs for an open system interacting with an environment \cite{Breuer2007,Joos2013}. Under the Born, Markov, and rotating-wave approximations, the dynamics of such a system is described by the Lindblad equation  \cite{Lindblad1976, Gorini1976, Breuer2007,Joos2013}. Here, we consider the following Lindblad equation 
to discuss the decoherence due to the loss of photons from the cavities,
\begin{equation}
\frac{d \rho }{dt} =- \frac{i}{\hbar} [ \hat{H}, \rho] 
+\kappa_\text{c} \sum^2_{i=1} \Bigr( \hat{c}_i \rho \hat{c}^\dagger_i 
- \frac{1}{2} \{ \hat{c}^\dagger_i \hat{c}_i, \rho \} \Bigr)
+\kappa_\text{c} \sum^2_{i=1} \Bigr( \hat{d}_i \rho \hat{d}^\dagger_i 
- \frac{1}{2} \{ \hat{d}^\dagger_i \hat{d}_i, \rho \} \Bigr), 
\label{eq:Lindblad1}
\end{equation} 
where 
$\{\hat{A} , \hat{B} \} = \hat{A}\hat{B}+\hat{B} \hat{A}$. We assumed the same damping constant 
$\kappa_\text{c}$ for each cavity along with an environment with zero temperature.
The second and third terms in the right hand side of Eq. (\ref{eq:Lindblad1}) 
correspond to the dissipation of photons into an environment. 
A similar model without gravity was investigated in \cite{Mancini1997}, 
and the effect of quantum decoherence for photons was discussed. In contrast, the decoherence for mechanical oscillators does not significantly affect the quantum state of photons because the effect is suppressed by the optomechanical couplings \cite{Bose1997}. 

One can solve Eq.~\eqref{eq:Lindblad1}
and find the reduced density operator of photons \eqref{eq:solLind}
after long but straightforward calculations.
Due to the non-unitary terms of the second term and the third term of 
the right hand side of Eq. \eqref{eq:Lindblad1}, $\rho_\text{c,d}(t)$ 
defined by \eqref{eq:rhocd} is modified as 
\begin{align}
\tilde \rho_\text{c,d}(t) 
&={\rm Tr}_\text{a,b}[\rho]
\nonumber
\\
&
= e^{-2\kappa_\text{c} t}\rho_\text{c,d}(t)
+ 
 (1-e^{-\kappa_\text{c} t})e^{-\kappa_\text{c} t} \, 
|0,0 \rangle_\text{c} \langle 0,0| \otimes \sigma_\text{d}
\nonumber \\
&
+ 
(1-e^{-\kappa_\text{c} t})e^{-\kappa_\text{c} t}
\sigma_\text{c} \otimes 
|0,0 \rangle_\text{d} \langle 0,0| 
+ 
(1-e^{-\kappa_\text{c}t})^2 \,   |0,0 \rangle_\text{c} \langle 0,0| 
\otimes 
|0,0 \rangle_\text{d} \langle 0,0|
\label{eq:solLind}
\end{align} 
where $\sigma_\text{c}$ and $\sigma_\text{d}$ are 
density operators with a single photon in the system c and d, respectively.
Their explicit expressions are presented in the Appendix C.
The last three terms in Eq.\eqref{eq:solLind} appear as the result of photon 
loss, which do not contribute quantum entanglement because they are written in 
the separable forms and have eigenstates orthogonal to those
of the first term with $\rho_\text{c,d}(t)$.
Hence the negativity described by Eq.~\eqref{eq:solLind} evolves as 
$e^{-2\kappa_\text{c} t} \mathcal{N}$ with the original negativity $\mathcal{N}$ given by Eq. \eqref{eq:explicitN}. 
This fact requires that the ratio 
$\kappa_\text{c}/\omega_\text{s}$ should be small to generate the entanglement of photons
sufficiently.

We can rewrite the condition, $\kappa_\text{c}/\omega_\text{s}<1$, 
in terms of the natural frequencies ($\Omega_\text{a}, \, \Omega_\text{b}$), 
the standard optomechanical couplings 
($\Lambda_\text{m}, \, \Lambda_\text{M}$ in Eq.~\eqref{eq:Lambda})
and the characteristic parameters 
($g_\text{a}, \, g_\text{b}$ in Eq.~\eqref{eq:g}) as 
\begin{equation}
\frac{\kappa_\text{c}}{\sqrt{\Omega_\text{a} \Omega_\text{b}}} < 
\frac{\sqrt{g_\text{a} g_\text{b}} \Lambda_\text{m} \Lambda_\text{M}} {1+g_\text{a} +g_\text{b}}.
\label{eq:ratio}
\end{equation}
The parameters 
$\Lambda_\text{m}$, $\Lambda_\text{M}$, 
$g_\text{a}$, and 
$g_\text{b}$ are assumed to be small so that the derivation of the Hamiltonian Eq.\eqref{eq:Htot} is valid. 
The right hand side of Eq.\eqref{eq:ratio} is $O(10^{-3})$
even for $\Lambda_\text{m} \sim \Lambda_\text{M} \sim O(0.1)$ and
$g_\text{a} \sim g_\text{b} \sim O(0.1)$.
It might be difficult to realize an experiment satisfying such a condition for the dissipation rate \cite{Aspelmeyer2014, Michimura2020}.
To store the gravity-induced entanglement in separated cavities, 
we require high-finesse optical cavities that satisfy Eq.\eqref{eq:ratio}.

\section{Conclusion}
\label{sec:4}

We investigated gravity-induced entanglement in optomechanical systems under a non-perturvative and nonlinear regime. Then, we clarified how the photons in separated cavities 
store the entanglement due to the Newtonian gravity between the mechanical 
oscillators. It is important to note that the photon entanglement
grows to reach 
the maximum value with a characteristic time $1/\omega_\text{s}$, defined by 
Eq.~\eqref{eq:slow}. 
The photon entanglement originates from the phase difference given by the potential minimum of the gravitating oscillators coupled with the photons in the cavities. 
The growth time of large photon entanglements (Eq.~\eqref{eq:slow}) is determined by the optomechcanical couplings and the gravitational interaction between the 
micro-mechanical oscillators. 

Since the growth time of the entanglement is relatively long, one requires the 
cavities confinig photons long time to generate large entanglements. 
However, quantum decoherence due to environmental interactions is inevitable and  a crucial issue 
for the generations of the entanglement. In our analysis with a simple model of photon 
leakage, we derived the condition for the dissipation constant
to store the photon entanglement. 
It will be challenging to realize optomechanical systems that can satisfy the required conditions 
in the present experimental technique. 

Another issue is the detection of quantum entanglement. 
The measurement of an entanglement witness plays an important role in characterizing the feasible 
detection of gravity-induced entanglement. Therefore, improved optomechanical cavities with a small dissipation rate and the detection of a suitable entanglement witness are required for testing quantum gravity. 

\begin{acknowledgments}
\end{acknowledgments}
We thank S. Kanno, J. Soda, Y. Nambu, N. Matsumoto, and Y. Michimura for useful discussions related to the topic of this paper. This work was supported by Ministry of Education, Culture, Sports, Science and Technology (MEXT)/Japan Society for the Promotion of Science (JSPS) KAKENHI Grant  No. 
17K05444 (KY).
\begin{appendix}

\section{Derivation of the total Hamiltonian of the optomechanical system}
\label{sec:A}

In this appendix, we review the derivation of the total Hamiltonian of the system (Eq. (\ref{eq:Htot})) presented in \cite{Balushi2018}. 
First, the Hamiltonian of the two mechanical rods is 
\begin{equation}
H_\text{m,M}
=
\frac{p^2_\text{m}}{2I_\text{m}} 
+ \frac{1}{2} I_\text{m} \Omega^2_\text{a} \theta^2_\text{m} 
+\frac{p^2_\text{M}}{2I_\text{M}} 
+ \frac{1}{2} I_\text{M} \Omega^2_\text{b} \theta^2_\text{M} + V_\text{g}, \label{eq:H}
\end{equation}
where  $I_m=2mL^2$ and $I_M =2ML^2$ are the two moments of inertia
and $V_\text{g}$ is the gravitational potential.  
Under the assumptions $|\theta_\text{m}|, |\theta_\text{M}| \ll 1$,  
$L^2/h^2 \times (\theta_\text{m} - \theta_\text{M})^2 \ll 1$ and 
$L/h \gg 1$, the gravitational potential is approximated as 
\begin{align}
V_\text{g} 
&=
-\frac{2GMm}
{\bigl[ h^2 + 2L^2 (1-\cos (\theta_\text{m} -\theta_\text{M})) \bigr]^{1/2}}
-\frac{2GMm}
{\bigl[ h^2 + 2L^2 (1+\cos (\theta_\text{m} -\theta_\text{M})) \bigr]^{1/2}}
\nonumber 
\\
&\simeq -\frac{2GMm}{h} \Bigl( \Bigl[ 1+ \frac{L^2}{h^2} (\theta_\text{m} - \theta_\text{M})^2 \Bigr]^{-1/2}+ \Bigl[ 1+ \frac{4L^2}{h^2} -\frac{L^2}{h^2} (\theta_\text{m} -\theta_\text{M})^2 \Bigr]^{-1/2} \Bigr)
\nonumber 
\\
&\simeq -\frac{2GMm}{h} \Bigl(  1- \frac{L^2}{2h^2} (\theta_\text{m} - \theta_\text{M})^2 + ( 1+4L^2/h^2 )^{-1/2} \Bigl[1+\frac{L^2/h^2}{2(1+4L^2/h^2)} (\theta_\text{m} -\theta_\text{M})^2 \Bigr] \Bigr)
\nonumber 
\\
&\simeq  -\frac{2GMm}{h} + \frac{GMmL^2}{h^3} (\theta_\text{m} -\theta_\text{M} )^2. \label{eq:Vg}
\end{align}
Here, we note that the gravitational potentials ($-Gm^2/L$ and $-GM^2/L^2$) between the masses of 
each rod as well as the uniform gravitational 
potential on the earth are constant; these are not relevant to the analysis
and are omitted. The term $-2GMm/h$ in the last line of Eq.~\eqref{eq:Vg} 
can be omitted. Then, the Hamiltonian of the two angular oscillators can be
rewritten as 
\begin{equation}
H_\text{m,M}
=\frac{p^2_\text{m}}{2I_\text{m}} 
+ \frac{1}{2} I_\text{m} \omega^2_\text{a} \theta^2_\text{m} 
+\frac{p^2_\text{M}}{2I_\text{M}} 
+ \frac{1}{2} I_\text{M} \omega^2_\text{b} \theta^2_\text{M} 
-\frac{2GMmL^2}{h^3} \theta_\text{m} \theta_\text{M}, \label{eq:H-2}
\end{equation}
where 
\begin{equation}
\omega^2_\text{a}=\Omega^2_\text{a} +\frac{GM}{h^3}, \quad 
\omega^2_\text{b}=\Omega^2_\text{b}+\frac{Gm}{h^3}. \label{eq:omegaab}
\end{equation}

Next, we review the optomechanical interaction between a photon and mirror 
(angular oscillator). The frequency of a photon inside a cavity of 
length $\ell$  is
\begin{equation}
\omega_c = 2\pi \frac{nc}{2\ell}=\frac{\pi n c}{\ell}, \label{eq:omegac}
\end{equation}
where 
$n=1,2,\dots$ and 
$c$ is the velocity of light. When the photon couples with the oscillating mirror attached to the edge of the mechanical rod, the frequency varies as 
\begin{equation}
\omega'_\text{c} = \frac{\pi n c}{\ell+L\sin \theta} \simeq \omega_\text{c} \Bigl(1-\frac{L}{\ell} \theta \Bigr). 
\end{equation}
where angle 
$\theta$ is assumed to be small. The Hamiltonian of the photon can be expressed as 
\begin{equation}
\hat{H}'_\text{c} = \hbar \omega'_\text{c} \hat{c}^\dagger \hat{c} \simeq \hbar \omega_\text{c} \hat{c}^\dagger \hat{c} -\frac{\hbar L\omega_c}{\ell} \hat{c}^\dagger \hat{c} \, \hat{\theta}. \label{eq:H'c}
\end{equation}
The last term of the right hand side of Eq.~(\ref{eq:H'c}) corresponds to the
optomechanical interaction. A further rigorous derivation and formulation of the optomechanical interaction was presented in Ref.~\cite{Law1995}. 

Finally, introducing the normalized variables
\begin{equation}
\hat{q}_\text{a}
=\sqrt{\frac{I_\text{m} \omega_\text{a}}{\hbar}} \hat{\theta}_\text{m}, 
\quad 
\hat{p}_\text{a}
=\frac{\hat{p}_\text{m}}{\sqrt{I_\text{m} \omega_\text{a}\hbar}},
\quad 
\hat{q}_\text{b}
=\sqrt{\frac{I_\text{M} \omega_\text{b}}{\hbar}} \hat{\theta}_\text{M}, 
\quad 
\hat{p}_\text{b}
=\frac{\hat{p}_\text{M}}{\sqrt{I_\text{M} \omega_\text{b}\hbar}}, \label{eq:ab}
\end{equation}
we obtain the total Hamiltonian of the oscillating mirrors and photons in the cavities as
\begin{align}
\hat{H}
&
= 
\hbar \omega_\text{c} 
(\hat{c}^\dagger_1 \hat{c}_1 + \hat{c}^\dagger_2 \hat{c}_2) 
+\hbar \omega_\text{d} 
(\hat{d}^\dagger_1 \hat{d}_1 + \hat{d}^\dagger_2 \hat{d}_2) 
-\lambda_\text{m} \hbar \omega_\text{a} \hat{c}^\dagger_1 \hat{c}_1 \, \hat{q}_\text{a}
-\lambda_\text{M} \hbar \omega_\text{b} \hat{d}^\dagger_1 \hat{d}_1 \, \hat{q}_\text{b} 
\nonumber 
\\
&
+\frac{\hbar \omega_\text{a}}{2} ( \hat{p}^2_\text{a}+\hat{q}^2_\text{a}) 
+\frac{\hbar \omega_\text{b}}{2} ( \hat{p}^2_\text{b}+\hat{q}^2_\text{b}) 
-\hbar \gamma \hat{q}_\text{a} \hat{q}_\text{b}, \label{eq:H1-2}
\end{align}
where 
\begin{align}
\lambda_\text{m} 
= \frac{\omega_\text{c}}{\omega_\text{a} \ell_\text{c}} \sqrt{\frac{\hbar}{2m \omega_\text{a}}}, 
\quad 
\lambda_\text{M} 
= \frac{\omega_\text{d}}{2\omega_\text{b} \ell_\text{d}} \sqrt{\frac{\hbar}{2M \omega_\text{b}}}, 
\quad 
\gamma
=\frac{G}{h^3} \sqrt{\frac{Mm}{\omega_\text{a} \omega_\text{b}}}. \label{eq:lambdagammaA}
\end{align}

\section{Computation of time evolution in the interaction picture}
\label{sec:B}

Here, we consider the operator of the time evolution in the interaction picture,
\begin{equation}
\text{T} \exp \Bigl[-\frac{i}{\hbar} \int^t_0 dt' \hat{V}_\text{I} (t') \Bigr],
\label{eq:TexpB_appendix}
\end{equation}
where the interaction potential is written as 
$\hat{V}_\text{I}(t) = -\hbar \, \hat{\bm{j}}^\text{T} \hat{\bm{\eta}}_\text{I}(t)$ with the photon operator, 
$\hat{\bm{j}}
= 
[\lambda_\text{m} \omega_\text{a} \hat{c}^\dagger_\text{1}\hat{c}_\text{1},  \lambda_\text{M} \omega_\text{b} \hat{d}^\dagger_\text{1}\hat{d}_\text{1}, 0, 0]^\text{T}$ and the canonical operator of the rods
$\hat{\bm{\eta}}_\text{I}(t)$ in the interaction picture. The solution 
$\hat{\bm{\eta}}_\text{I}(t)$ is given by the linear combination of 
$\hat{\bm{\eta}} 
= 
[ \hat{q}_\text{a}, \hat{q}_\text{b}, \hat{p}_\text{a}, \hat{p}_\text{b} ]^{\text{T}}$ as
\begin{equation}
\hat{\bm{\eta}}_\text{I}(t)=e^{t\Omega \mathcal{H}} \hat{\bm{\eta}}.
\label{eq:soletaB}
\end{equation}
The interaction $\hat{V}_\text{I}$ does not change the photon number : 
$[\hat{V}_\text{I}(t), \hat{c}^\dagger_1 \hat{c}_1]=[\hat{V}_\text{I}(t), \hat{d}^\dagger_1 \hat{d}_1]=0$, which leads to 
the following important equation;
\begin{equation}
[ \hat{V}_\text{I} (t_1), [ \hat{V}_\text{I} (t_2), \hat{V}_\text{I} (t_3)]]=0.
\label{eq:VVV}
\end{equation}
Using Eq.\eqref{eq:VVV}, we explicitly compute the operator of time evolution (Eq. \eqref{eq:TexpB_appendix}) as 
\begin{align}
&
\text{T} \exp \Bigl[-\frac{i}{\hbar} \int^t_{0} dt' \hat{V}_\text{I} (t') \Bigr]
\nonumber
\\
&
\quad 
\quad
=\lim_{N\rightarrow \infty } 
\exp \Bigl[-\frac{i}{\hbar} \Delta t \hat{V}_\text{I} (t_{N-1}) \Bigr] 
\cdots
\exp \Bigl[-\frac{i}{\hbar} \Delta t \hat{V}_\text{I} (t_{1}) \Bigr] \exp \Bigl[-\frac{i}{\hbar} \Delta t \hat{V}_\text{I} (t_0) \Bigr] 
\nonumber
\\
&
\qquad
=\lim_{N\rightarrow \infty } 
\exp \Bigl[-\frac{i}{\hbar} \Delta t \hat{V}_\text{I} (t_{N-1}) \Bigr] 
\cdots
\exp \Bigl[-\frac{i}{\hbar} \Delta t \hat{V}_\text{I} (t_{2}) \Bigr]  
\nonumber 
\\
&
\qquad
\quad
\times
\exp 
\Bigl[
-\frac{i}{\hbar} \Delta t \bigl(\hat{V}_\text{I} (t_{1})+\hat{V}_\text{I} (t_0) \bigr)  
-\frac{\Delta t^2}{\hbar^2} [\hat{V}_\text{I} (t_{1}), \hat{V}_\text{I} (t_0)]
\Bigr] 
\nonumber
\\
&
\qquad
=\lim_{N\rightarrow \infty } 
\exp \Bigl[-\frac{i}{\hbar} \Delta t \hat{V}_\text{I} (t_{N-1}) \Bigr] 
\cdots
\exp \Bigl[-\frac{i}{\hbar} \Delta t \hat{V}_\text{I} (t_{3}) \Bigr] 
\nonumber
\\
&
\qquad
\quad
\times
\exp 
\Bigl[
-\frac{i}{\hbar} \Delta t 
\bigl(
\hat{V}_\text{I} (t_{2})+\hat{V}_\text{I} (t_{1})+\hat{V}_\text{I} (t_0) 
\bigr)  
-\frac{\Delta t^2}{\hbar^2}  
\bigl( 
[\hat{V}_\text{I} (t_{2}), \hat{V}_\text{I} (t_{1})+\hat{V}_\text{I} (t_0) ]
+
[\hat{V}_\text{I} (t_{1}), \hat{V}_\text{I} (t_0)] 
\bigr)
\Bigr] 
\nonumber
\\
&
\qquad 
=\lim_{N\rightarrow \infty } 
\exp 
\Bigl[
-\frac{i}{\hbar} \Delta t \sum^{N-1}_{i=0} \hat{V}_\text{I} (t_{i}) 
-\frac{\Delta t^2}{2\hbar^2}  \sum^{N-1}_{i>j=0} 
[\hat{V}_\text{I} (t_i), \hat{V}_\text{I} (t_j)]
 \Bigr] 
\nonumber 
\\
&
\qquad
=
\exp 
\Bigl[
-\frac{i}{\hbar} \int^t_0 dt'  \hat{V}_\text{I} (t') 
-\frac{1}{2\hbar^2} \int^t_0 dt' \int^{t'}_0 dt''  
[\hat{V}_\text{I} (t'), \hat{V}_\text{I} (t'')]
 \Bigr] ,
\label{eq:TexpB2_appendix}
\end{align}
where 
$\Delta t = (t-t_0)/N$ and 
$t_0 =0$. In the derivation of Eq.~\eqref{eq:TexpB2_appendix}, we sequentially used the 
Backer-Campbell-Hausdorff formula and Eq.~\eqref{eq:VVV} as follows:
\begin{align}
&
\exp \Bigl[-\frac{i}{\hbar} \Delta t \hat{V}_\text{I} (t_i) \Bigr] 
\exp \Bigl[-\frac{i}{\hbar} \Delta t \hat{V}_\text{I} (t_j) \Bigr] 
\nonumber
\\
&
=
\exp 
\Bigl[
-\frac{i}{\hbar} \Delta t \bigl( \hat{V}_\text{I} (t_i) +\hat{V}_\text{I} (t_j) \bigr)
-\frac{\Delta t^2}{2\hbar^2}  [\hat{V}_\text{I} (t_i), \hat{V}_\text{I} (t_j)] 
+\frac{i\Delta t^3}{12\hbar^3} [\hat{V}_\text{I} (t_i) -\hat{V}_\text{I} (t_j), [\hat{V}_\text{I} (t_i), \hat{V}_\text{I} (t_j)] ]+ \cdots 
 \Bigr] 
\nonumber 
\\
&
=
\exp 
\Bigl[
-\frac{i}{\hbar} \Delta t \bigl( \hat{V}_\text{I} (t_i) +\hat{V}_\text{I} (t_j) \bigr)
-\frac{1}{2\hbar^2}  \Delta t^2 [\hat{V}_\text{I} (t_i), \hat{V}_\text{I} (t_j)] 
 \Bigr].
\label{eq:BCH}
\end{align}
Substituting 
$\hat{V}_\text{I}(t) =- \hbar \, \hat{\bm{j}}^\text{T} \hat{\bm{\eta}}_\text{I}(t)$ into Eq.~\eqref{eq:TexpB2_appendix}, we obtain 
\begin{align}
\text{T} \exp \Bigl[-\frac{i}{\hbar} \int^t_0 dt' \hat{V}_\text{I} (t') \Bigr]
&
=
\exp 
\Bigl[
-\frac{i}{\hbar} \int^t_0 dt'  \hat{V}_\text{I} (t') 
-\frac{1}{2\hbar^2} \int^t_0 dt' \int^{t'}_0 dt''  
[\hat{V}_\text{I} (t'), \hat{V}_\text{I} (t'')]
 \Bigr]
\nonumber
\\
&
=
\exp 
\Bigl[
i \int^t_0 dt'  \, \hat{\bm{j}}^\text{T} \hat{\bm{\eta}}_\text{I} (t')  
-\frac{1}{2} \int^t_0 dt' \int^{t'}_0 dt'' \, \sum_{i,k} \hat{j}^i
[\hat{\eta}^i_\text{I} (t'),  \hat{\eta}^k_\text{I} (t'')] \hat{j}^k 
 \Bigr]
\nonumber
\\
&
=
\exp 
\Bigl[
i \int^t_0 dt'  \, \hat{\bm{j}}^\text{T} \hat{\bm{\eta}}_\text{I} (t')  
-\frac{i}{2} \int^t_0 dt' \int^{t}_0 dt'' \, 
\hat{\bm{j}}^\text{T} {\cal G}(t',t'') \hat{\bm{j}} 
 \Bigr],
\label{eq:TexpB2}
\end{align}
where ${\cal G}(t',t'') $ denotes the matrix of the retarded Green function 
with the component 
${\cal G}^{jk}(t',t'')
= -i [\hat{\eta}^j_\text{I} (t'),  \hat{\eta}^k_\text{I} (t'')] \theta(t' -t'')$.

\section{Solution of the Lindblad equation \eqref{eq:Lindblad1}}
\label{sec:C}

We present the solution of Eq.\eqref{eq:Lindblad1} 
and give the formulas for
the density operators 
$\sigma_\text{c}$
and
$\sigma_\text{d}$ in Eq.\eqref{eq:solLind}. As the first step, we consider the following Lindblad equation,
\begin{equation}
\frac{d \rho }{dt} =- \frac{i}{\hbar} [ \hat{K}, \rho] 
+\kappa  \Bigl( \hat{a} \rho \hat{a}^\dagger 
- \frac{1}{2} \{ \hat{a}^\dagger \hat{a}, \rho \} \Bigr), 
\label{eq:Lindblad2}
\end{equation}
where 
$\hat{K}$ 
is a Hamiltonian, 
$\hat{a} \, (\hat{a}^\dagger)$ is an annihilation (creation) operator satisfying 
$[\hat{a},\hat{a}^\dagger]=1$ and
$\kappa$ is a dissipation rate. 
We assume that the number operator 
$\hat{a}^\dagger \hat{a}$ and
the Hamiltonian 
$\hat{K}$ commute each other, $[\hat{K}, \hat{a}^\dagger \hat{a}]=0$. We define the density operator 
$\rho'
=
e^{i\hat{K}t/\hbar} 
\rho e^{-i\hat{K}t/\hbar}$ and then find the equation 
\begin{equation}
\frac{d \rho' }{dt} =
\kappa  \Bigl( \hat{A}(t) \rho' \hat{A}^\dagger (t) 
- \frac{1}{2} \{ \hat{A}^\dagger(t) \hat{A}(t), \rho' \} \Bigr)
=
\kappa  \Bigl( \hat{A}(t) \rho' \hat{A}^\dagger (t) 
- \frac{1}{2} \{ \hat{a}^\dagger \hat{a}, \rho' \} \Bigr), 
\label{eq:Lindblad3}
\end{equation}
where we defined 
$\hat{A}(t)=e^{i\hat{K}t/\hbar} \hat{a} e^{-i\hat{K}t/\hbar}$, and 
used the fact that the number operator commutes with the Hamiltonian 
$\hat{K}$ in the second equality. Introducing the operator 
$\tilde{\rho}
=
e^{\kappa t \hat{a}^\dagger \hat{a}/2}
\rho' e^{\kappa t \hat{a}^\dagger \hat{a}/2}$, we obtain the equation 
for
$\tilde{\rho}$ as follows: 
\begin{equation}
\frac{d \tilde{\rho} }{dt} 
=
\kappa  e^{\kappa t \hat{a}^\dagger \hat{a}/2} \hat{A}(t) 
e^{-\kappa t \hat{a}^\dagger \hat{a}/2} \tilde{\rho} e^{-\kappa t \hat{a}^\dagger \hat{a}/2}
\hat{A}^\dagger (t) 
e^{\kappa t \hat{a}^\dagger \hat{a}/2}
=
\kappa e^{-\kappa t}  \hat{A}(t) 
 \tilde{\rho} 
\hat{A}^\dagger (t),
\label{eq:Lindblad4}
\end{equation}
where in the second equality we used that 
$ e^{\kappa t \hat{a}^\dagger \hat{a}/2} 
\hat{A}(t) 
e^{-\kappa t \hat{a}^\dagger \hat{a}/2}
=
e^{i\hat{K}t/\hbar} 
e^{\kappa t \hat{a}^\dagger \hat{a}/2} \hat{a} 
e^{-\kappa t \hat{a}^\dagger \hat{a}/2}
e^{-i\hat{K}t/\hbar}
=
e^{-\kappa t/2} \hat{A}(t)$ holds by 
$[\hat{K}, \hat{a}^\dagger \hat{a}]=0$ and the Baker-Campbell-Hausdorff formula. The formal solution of Eq. \eqref{eq:Lindblad4} is given as 
\begin{equation}
\tilde{\rho}(t) =  \sum^{\infty}_{m=0}
\int^t_0 dt_1 \int^{t_1}_0 dt_2 \cdots
\int^{t_{m-1}}_0 dt_{m} \, \mathcal{J}_{t_1}
[\mathcal{J}_{t_2}
[\cdots[
\mathcal{J}_{t_m}
[ \tilde{\rho}(0)
] \cdots]=
\text{T} \exp 
\Bigl[ 
\int^t_0 dt' \mathcal{J}_{t'}
\Bigr][\tilde{\rho}(0)],
\label{eq:Lindblad4}
\end{equation}
where 
$\mathcal{J}_t$ is the superoperator defined by  $\mathcal{J}_t[\hat{O}]
=
\kappa e^{-\kappa t} 
\hat{A}(t) 
\hat{O} 
\hat{A}^\dagger(t)$ and 
T is the time ordering. Hence the solution with an initial state 
$\rho(0)$ of the Lindblad equation \eqref{eq:Lindblad3} is obtained as \cite{Walls1985} 
\begin{equation}
\rho(t)
= e^{-i\hat{K} t/\hbar -\kappa t \hat{a}^\dagger \hat{a}/2}  
\,
\text{T} \exp 
\Bigl[ 
\int^t_0 dt' \mathcal{J}_{t'}
\Bigr][\rho(0)]
\,
e^{i\hat{K} t/\hbar -\kappa t \hat{a}^\dagger \hat{a}/2}.
\label{eq:solLin2}
\end{equation}

Since the total Hamiltonian \eqref{eq:Htot} commutes with the photon number operators, we can apply the above procedure to derive the solution of the Lindblad equation \eqref{eq:Lindblad1}. The solution is given by 
\begin{equation}
\rho(t)
= e^{-i\hat{H} t/\hbar -\frac{\kappa_\text{c} t}{2} \sum_{i=1,2}(\hat{c}^\dagger_i \hat{c}_i + \hat{d}^\dagger_i \hat{d}_i) }  
\,
\text{T} \exp 
\Bigl[ 
\int^t_0 dt' \mathcal{S}_{t'}
\Bigr][\rho(0)]
\,
e^{i\hat{H} t/\hbar -\frac{\kappa_\text{c} t}{2} \sum_{i=1,2}(\hat{c}^\dagger_i \hat{c}_i + \hat{d}^\dagger_i \hat{d}_i) },
\label{eq:solLin3}
\end{equation}
where the superoperator 
$\mathcal{S}_t$ is deifned as
\begin{equation}
\mathcal{S}_t[\hat{O}]
=\kappa_\text{c} e^{-\kappa_\text{c} t} 
\sum_{i=1,2}
\Bigl( 
\hat{C}_i (t) 
\hat{O} 
\hat{C}^\dagger_i (t) 
+
\hat{D}_i (t) 
\hat{O} 
\hat{D}^\dagger_i (t) 
\Bigr),
\label{eq:S}
\end{equation}
with 
$\hat{C}_i (t) =e^{i\hat{H}t/\hbar} \hat{c}_i e^{-i\hat{H}t/\hbar}$ and
$\hat{D}_i (t) =e^{i\hat{H}t/\hbar} \hat{d}_i e^{-i\hat{H}t/\hbar}$. From the formula \eqref{eq:solLin3}, we can compute the solution \eqref{eq:solLind},
\begin{align}
\tilde \rho_\text{c,d}(t)
&
= e^{-2\kappa_\text{c} t}\rho_\text{c,d}(t)
+ 
 (1-e^{-\kappa_\text{c} t})e^{-\kappa_\text{c} t} \, 
|0,0 \rangle_\text{c} \langle 0,0| \otimes \sigma_\text{d}
\nonumber \\
&
+ 
(1-e^{-\kappa_\text{c} t})e^{-\kappa_\text{c} t}
\sigma_\text{c} \otimes 
|0,0 \rangle_\text{d} \langle 0,0| 
+ 
(1-e^{-\kappa_\text{c}t})^2 \,   |0,0 \rangle_\text{c} \langle 0,0| 
\otimes 
|0,0 \rangle_\text{d} \langle 0,0|, 
\label{eq:solLin5}
\end{align}
where the initial condition 
$\rho(0)$ is
$ |\phi \rangle_\text{c,d} 
\langle \phi| 
\otimes |\bm{\xi} \rangle_\text{a,b} 
\langle \bm{\xi}|$
with 
$|\phi \rangle_\text{c,d}$ and 
$|\bm{\xi} \rangle_\text{a,b}$ defined by 
\eqref{eq:inphoton} and \eqref{eq:xi}, respectively. The density operators 
$\sigma_\text{c}$ and  $\sigma_\text{d}$ are 
\begin{align}
\sigma_\text{c}
&
= 
\frac{1}{4} 
\int^t_0 dt' \frac{\kappa_\text{c}e^{-\kappa_\text{c} t'}}
{1-e^{-\kappa_\text{c} t}} \sum_{m=0,1} \sum_{n,n'=0,1}
\langle \Phi^{n' 0}_{n' m}(t,t')|\Phi^{n 0}_{n m} (t,t') \rangle
|n,1-n \rangle_\text{c} 
\langle n',1-n' |,
\label{eq:sigmac}
\\
\sigma_\text{d}
&
= 
\frac{1}{4} 
\int^t_0 dt' \frac{\kappa_\text{c}e^{-\kappa_\text{c} t'}}
{1-e^{-\kappa_\text{c} t}} \sum_{m,m'=0,1} \sum_{n=0,1}
\langle \Phi^{0 m'}_{n m'}(t,t')|\Phi^{0 m}_{n m} (t,t') \rangle
|m,1-m \rangle_\text{d} 
\langle m',1-m' |,
\label{eq:sigmad}
\end{align}
where 
\begin{equation}
|\Phi^{n'm'}_{nm}(t,t') \rangle
=
e^{-i (t-t') ( \hat{\bm{\eta}}^\text{T} \mathcal{H} \hat{\bm{\eta}}/2 - \bm{j}^\text{T}_{n'm'} \hat{\bm{\eta}}) }  e^{-i t' ( \hat{\bm{\eta}}^\text{T} \mathcal{H} \hat{\bm{\eta}}/2 - \bm{j}^\text{T}_{nm} \hat{\bm{\eta}}) }
| \bm{\xi} \rangle_\text{a,b}.
\label{eq:Phi}
\end{equation}
Here,  
$\hat{\bm{\eta}}$ and 
$\mathcal{H}$ are defined in Eq.\eqref{eq:etah} and Eq.\eqref{eq:mathH}, and $\bm{j}_{nm}=[\lambda_\text{m} \omega_\text{a} n, \lambda_\text{M} \omega_\text{b}m,0,0]^\text{T}$. 

\end{appendix}



\begin{thebibliography}{10}
\newcommand{\enquote}[1]{``#1''}

\bibitem{Kiefer2006}
C. Kiefer, \enquote{Quantum gravity: general introduction and recent developments}, \emph{Ann. Phys.} \textbf{15}, 129 (2006)

\bibitem{Woodard2009}
R. P. Woodard, \enquote{How far are we from the quantum theory of gravity?}, \emph{Rep. Prog. Phys.} \textbf{72}, 126002 (2009)

\bibitem{Schmole2016}
J. Schm\"{o}le, M. Dragosits, H. Hepach, and M. Aspelmeyer, \enquote{A micromechanical proof-of-principle experiment for measuring the gravitational force of milligram masses} \emph{Class. Quantum Grav.} \textbf{33}, 125031 (2016)
 
\bibitem{Matsumoto2019}
N. Matsumoto, S. B. Cata$\tilde{\text{n}}$o-Lopez, M. Sugawara, S. Suzuki, N. Abe, K. Komori, Y. Michimura, Y. Aso, and K. Edamatsu, \enquote{Demonstration of Displacement Sensing of a mg-Scale Pendulum for mm- and mg-Scale Gravity Measurements} \emph{Phys. Rev. Lett.} \textbf{122}, 071101 (2019)

\bibitem{Matsumoto2020}
S. B. Cata$\tilde{\text{n}}$o-Lopez, J. G. Santiago-Condori, K. Edamatsu, and N. Matsumoto, \enquote{High-$Q$ Milligram-Scale Monolithic Pendulum for Quantum-Limited Gravity Measurements},\emph{Phys. Rev. Lett.} \textbf{124}, 221102 (2020)
 
\bibitem{Pikovski2012}
I. Pikovski, M. R. Vanner, M. Aspelmeyer, M. S. Kim, and $\check{\text{C}}$. Brukner, \enquote{Probing Planck-scale physics with quantum optics},  \emph{Nat. Phys.} \textbf{8}, 393 (2012).

\bibitem{Albrecht2014}
A. Albrecht, A. Retzker, and M. B. Plenio, \enquote{Testing quantum gravity by nanodiamond interferometry with nitrogen-vacancy centers}, \emph{Phys. Rev. A} \textbf{90}, (2014) 033834.

\bibitem{Grossardt2016}
A. Gro{\ss}ardt, J. Bateman, H. Ulbricht, and A. Bassi \enquote{Optomechanical test of the Schrödinger-Newton equation}, \emph{Phys. Rev. D} \textbf{93}, (2016) 096003.

\bibitem{Bose2017}
S. Bose, A. Mazumdar, G. W. Morley, H. Ulbricht, M Toro$\check{\text{s}}$, M. Paternostro, A. A. Geraci, P. F. Barker, M. S. Kim, and G. Milburn, \enquote{Spin Entanglement Witness for Quantum Gravity}, \emph{Phys. Rev. Lett.} \textbf{119}, (2017) 240401.

\bibitem{Marletto2017}
C. Marletto and V. Vedral, \enquote{Gravitationally Induced Entanglement between Two Massive Particles is Sufficient Evidence of Quantum Effects in Gravity}, \emph{Phys. Rev. Lett.} \textbf{119}, (2017) 240402.

\bibitem{Balushi2018}
A. A. Balushi, W. Cong, and R. B. Mann, \enquote{Optomechanical quantum Cavendish experiment}, \emph{Phys. Rev. A} \textbf{98}, (2018) 043811.

\bibitem{Miao2020}
H. Miao, D. Martynov, H. Yang, and A. Datta, \enquote{Quantum correlations of light mediated by gravity}, \emph{Phys. Rev. A} \textbf{101}, (2020) 063804.

\bibitem{CWan}
C. Wan, \enquote{Quantum superposition on nano-mechanical oscillator}, PhD thesis (Imperial College London, 2017)

\bibitem{Hall2018}
M. J. Hall and M. Reginatto, \enquote{On two recent proposals for witnessing nonclassical gravity}, \emph{J. Phys. A} \textbf{51}, 085303 (2018).

\bibitem{Belenchia2018}
A. Belenchia, R. M. Wald, F. Giacomini, E. Castro-Ruiz, $\check{\text{C}}$. Brukner, and M. Aspelmeyer, \enquote{Quantum superposition of massive objects and the quantization of gravity}, \emph{Phys. Rev.D} \textbf{98}, (2018) 126009.

\bibitem{Carney2019}
D. Carney, P. C. E. Stamp, and J. M. Taylor, \enquote{Tabletop experiments for quantum gravity: a users manual}, \emph{Class. Quant. Grav.} \textbf{36}, (2019) no. 3, 034001.

\bibitem{Carlesso2019}
M. Carlesso, A. Bassi, M. Paternostro, and H. Ulricht, \enquote{Testing the gravitational field generated by a quantum superposition}, \emph{New. J. Phys.} \textbf{21}, (2019) 093052.

\bibitem{Marshman2020}
R. J. Marshman, A. Mazumdar, and S.Bose, \enquote{Locality and entanglement in table-top testing of the quantum nature of linearized gravity}, \emph{Phys. Rev. A} \textbf{101}, (2020) 052110.

\bibitem{Krisnanda2020}
T. Krisnanda, G. Y. Tham, M. Paternostro, and T. Paterek, \enquote{Observable quantum entanglement due to gravity}, \emph{npj Quantum Inf.} \textbf{6}, 12 (2020).

\bibitem{Nielsen2002}
M. A. Nielsen and I. Chuang, \enquote{Quantum Computation and Quantum Information} (Cambridge University Press, Cambridge, England, 2002).

\bibitem{Horodecki2009}
R. Horodecki, P. Horodecki, M. Horodecki, and K. Horodecki, \enquote{Quantum entanglement}, \emph{Rev. Mod. Phys.} \textbf{81}, (2009) 865.

\bibitem{Chen2013}
Y. Chen, \enquote{Macroscopic Quantum Mechanics: Theory and Experimental Concepts of Optomechanics}, \emph{J. Phys. B} \textbf{46}, 104001 (2013).

\bibitem{Aspelmeyer2014}
M. Aspelmeyer, T. J. Kippenberg, and F. Marquardt, \enquote{Cavity optomechanics}, \emph{Rev. Mod. Phys. } \textbf{86}, (2014) 1391.

\bibitem{Marshall2003}
W. Marshall, C. Simon, R. Penrose, and D. Bouwmeester, \enquote{Towards Quantum Superpositions of a Mirror}, \emph{Phys. Rev. Lett.} \textbf{91}, (2003) 130401.

\bibitem{Adler2005}
S. L. Adler, A. Bassi, and E. Ippoliti, \enquote{Towards quantum superpositions of a mirror: an exact open systems analysis—calculational details}, \emph{J. Phys. A: Math. Gen.} \textbf{38},2715 (2005).

\bibitem{Kleckner2008}
D. Kleckner, I. Pikovski, E. Jeffrey, L. Ament, E. Eliel, J. Van Den Brink, and D. Bouwmeester, \enquote{Creating and verifying a quantum superposition in a micro-optomechanical system} \emph{New J. Phys.} \textbf{10}, 095020 (2008).

\bibitem{Mancini1997}
S. Mancini, V. I. Man'ko, and P. Tombesi, \enquote{Ponderomotive control of quantum macroscopic coherence}, \emph{Phys. Rev. A} \textbf{55}, (1997) 3042.

\bibitem{Bose1997}
S. Bose, K. Jacobs, and P. L. Knight, \enquote{Preparation of nonclassical states in cavities with a moving mirror}, \emph{Phys. Rev. A} \textbf{56}, (1997) 4175.

\bibitem{Law1995}
C. K. Law, \enquote{Interaction between a moving mirror and radiation pressure: A Hamiltonian formulation}, \emph{Phys. Rev. A} \textbf{51}, (1995) 2537.

\bibitem{Vidal2002}
G. Vidal and R. Werner, \enquote{Computable measure of entanglement}, \emph{Phys. Rev. A} \textbf{65}, (2002) 032314.

\bibitem{Horodecki1996}
M. Horodecki, R. Horodecki, and P. Horodecki, \enquote{Separability of mixed states: necessary and sufficient conditions}, \emph{Phys. Lett. A} \textbf{223}, (1996) 1-8.

\bibitem{Peres1996}
A. Peres,  \enquote{Separability Criterion for Density Matrices}, \emph{Phys. Rev. Lett.} \textbf{77}, (1996) 1413.

\bibitem{Breuer2007} 
H. P. Breuer and F. Petruccione, \enquote{The Theory of Open Quantum Systems},  (Oxford University Press, Oxford, 2007)

\bibitem{Joos2013} 
E. Joos, H. D. Zeh, C. Kiefer, D. J. W. Giulini, J. Kupsch, and I.-O. Stamatescu, \enquote{Decoherence and the Appearance of a Classical World in Quantum Theory}, (Springer-Verlag, Berlin Heidelberg, 2013)
  
\bibitem{Lindblad1976} 
G. Lindblad,  \enquote{On the generators of quantum dynamical semigroups}. \emph{Commun. Math. Phys.} \textbf{48} 119 (1976).

\bibitem{Gorini1976} 
V. Gorini, A. Kossakowski, E.C.G. Sudarshan, \enquote{Completely positive dynamical semigroups of N-level systems}, \emph{J. Math. Phys.} \textbf{17}, 821 (1976).

\bibitem{Walls1985}
D. F. Walls and G. J. Milburn, \enquote{Effect of dissipation on quantum coherence}, \emph{Phys. Rev. A} \textbf{31}, (1985) 2403.

\bibitem{Michimura2020}
Y. Michimura and K. Komori, \enquote{Quantum sensing with milligram scale optomechanical systems}, \emph{Eur. Phys. J. D} (2020) \textbf{73}, 125. 





\end{thebibliography}
\end{document}